\documentclass[useAMS,usenatbib]{mn2e}

\usepackage{graphicx,color}

\title[Solar twins in the open cluster M67]
{The chemical compositions of solar twins in the open cluster M67\thanks{The data presented herein were obtained at the W.M.\ Keck Observatory, which is operated as a scientific partnership among the California Institute of Technology, the University of California and the National Aeronautics and Space Administration. The Observatory was made possible by the generous financial support of the W.M.\ Keck Foundation.}}

\author[F. Liu et al.]{F. Liu$^{1,5}$\thanks{E-mail: fan.liu@astro.lu.se},
M. Asplund$^{1}$,
D. Yong$^{1}$,
J. Mel\'endez$^{2}$,
I. Ram\'irez$^{3}$,
A. I. Karakas$^{1,4}$,
\newauthor
M. Carlos$^{2}$,
A. F. Marino$^{1}$\\
$^{1}$Research School of Astronomy and Astrophysics, Australian National University, Canberra, ACT 2611, Australia\\
$^{2}$Departamento de Astronomia do IAG/USP, Universidade de Sao Paulo, Rua do Matao 1226, Sao Paulo 05508-900, SP, Brasil\\
$^{3}$McDonald Observatory and Department of Astronomy, University of Texas at Austin, 2515 Speedway, Austin, TX 78712-1205, USA\\
$^{4}$Monash Centre for Astrophysics, School of Physics and Astronomy, Monash University, VIC 3800, Australia\\
$^{5}$Lund Observatory, Department of Astronomy and Theoretical physics, Lund University, Box 43, SE-22100 Lund, Sweden}

\begin{document}

\date{Accepted ? Received ?; in original form 2016 April 11}

\pagerange{\pageref{firstpage}--\pageref{lastpage}} \pubyear{2016}

\maketitle

\label{firstpage}

\begin{abstract}
Stars in open clusters are expected to share an identical abundance pattern. Establishing the level of chemical homogeneity in a given open cluster deserves further study as it is the basis of the concept of chemical tagging to unravel the history of the Milky Way. 
M67 is particularly interesting given its solar metallicity and age as well as being a dense cluster environment. We conducted a strictly line-by-line differential chemical abundance analysis of two solar twins in M67: M67-1194 and M67-1315. Stellar atmospheric parameters and elemental abundances were obtained with high precision using Keck/HIRES spectra. M67-1194 is essentially identical to the Sun in terms of its stellar parameters. M67-1315 is warmer than M67-1194 by $\approx$ 150 K as well as slightly more metal-poor than M67-1194 by $\approx$ 0.05 dex. M67-1194 is also found to have identical chemical composition to the Sun, confirming its solar twin nature. The abundance ratios [X/Fe] of M67-1315 are similar to the solar abundances for elements with atomic number Z $\le$ 30, while most neutron-capture elements are enriched by $\approx$ 0.05 dex, which might be attributed to enrichment from a mixture of AGB ejecta and $r$-process material. The distinct chemical abundances for the neutron-capture elements in M67-1315 and the lower metallicity of this star compared to M67-1194, indicate that the stars in M67 are likely not chemically homogeneous. This poses a challenge for the concept of chemical tagging since it is based on the assumption of stars forming in the same star-forming aggregate. 
\end{abstract}


\section{Introduction}

The technique of a strictly line-by-line differential analysis in relative chemical abundances for similar type stars can yield extremely high precision (0.01 - 0.02 dex). This has recently been applied to various cases (see e.g.,  \citealp{mel09,mel12,yon13,liu14,liu16a,liu16b,ram14,ram15,tm14,bia15,nis15,saf15,spi16}). Among other things this has revealed subtle chemical abundance differences in the photospheres of stars, which has been interpreted as a signature of planet formation. \citet{mel09} discovered that the Sun exhibits a peculiar chemical pattern when compared to the majority of solar twins, namely, a depletion of refractory elements relative to volatile elements. They attributed this abundance pattern to the formation of terrestrial planets in the solar system, leading to a deficiency of refractory elements in the solar photosphere. In their scenario, refractory elements in the proto-solar nebula were locked up in the terrestrial planets. The remaining dust-cleansed gas was then accreted onto the Sun. In contrast, it is assumed that a typical solar twin did not form remaining terrestrial planets as efficiently. Alternatively, the dust-cleansed accretion happened so early that stellar convection was deep enough to dilute the refractory-poor gas sufficiently and thereby not leaving a detectable signature in the photosphere abundances. \citet{cha10} confirmed quantitatively that the depletion of refractories in the solar photosphere is consistent with the removal of $\approx$ 4 Earth masses of rocky material from otherwise pristine composition (see also Fig. 8 in \citealp{yg16a}).

The planet formation scenario, however, has been challenged by \citet{gh10} and \citet{adi14}. They argued that the observed trend between elemental abundances and dust condensation temperature (T$_{\rm cond}$) could instead be due to the differences in stellar ages rather than planet formation. \citet{nis15} showed clear abundance-age correlations for 21 solar-twins, which indicates that chemical evolution in the Galactic disc may play an important role in the explanation of the T$_{\rm cond}$ trends, although still a genuine dependence on T$_{\rm cond}$ seems to remain; this has subsequently been confirmed and extended by \citet{spi16} and \citet{nis16}.

Another possible explanation for the peculiar solar composition is that some of the dust in the pre-solar nebula was radiatively cleansed by luminous hot stars in the solar neighbourhood before the formation of the Sun and its planets. This dust-cleansing scenario is supported by the finding that the open cluster M67 seems to have a chemical composition closer to the solar composition than most solar twins (\citealp{one11}, hereafter O11 and \citealp{one14}, hereafter O14). They suggested that the proto-solar nebula was dust-cleansed by massive stars, similar to what happened for the proto-cluster cloud of M67, while the majority of solar twins in the field would presumably have formed in less massive clusters where no nearby high-mass star ($\ge$ 15 M$_\odot$) was formed. M67 offers the possibility for studying the solar-type stars in a dense cluster environment. This cluster has about solar metallicity ([Fe/H] in the range $-$0.04 to $+$0.03, e.g., \citealp{ht91,yon05,ran06,pas08}). The age of M67 is also comparable with that of the Sun: 3.5 - 4.8 Gyr \citep{yad08}. \citet{pas08} listed 10 solar-twin candidates in M67, including M67-1194 which hosts a hot Jupiter with a period of 6.9 days and a minimum mass of 0.34 M$_{\rm Jup}$ \citep{bru14}. M67-1194 was studied by O11 and re-visited by O14. They found that unlike nearby solar twins, which were systematically analyzed by \citet{mel09} and \citet{ram09}, the chemical composition of M67-1194 is more solar-like and therefore suggested that the Sun may have been formed in a similar cluster environment, perhaps even in M67. While this scenario is plausible when considering chemical abundances and ages, the dynamics are problematic: the Sun has an orbit close to the Galactic plane while M67 is presently some 450 pc above the plane. The probability that the Sun, if formed in M67 with an orbit similar to the present cluster orbit, was scattered or diffused out of the cluster into the present solar orbit, is found to be quite low, if the existence of the outer planetary system is taken into account \citep{pic12}. \citet{gus16} suggested that the high-altitude metal-rich clusters (such as M67) were formed in orbits close to the Galactic plane and later scattered to higher orbits by interaction with giant molecular clouds and spiral arms. Thus it is possible, though not very probable, that the Sun formed in such a cluster before scattering occurred. Currently only one solar-twin in M67 (M67-1194) was spectroscopically analyzed with high precision ($\sim$ 0.02 dex). It is thus crucial to analyze the chemical composition of additional solar-type stars in M67 of very high precision.

Stellar chemical abundances are expected to keep the fossil record of the conditions of the Galaxy at the time of its formation, at least for dwarf stars\footnote{Stellar depletion of Li and Be are obvious exceptions. Similarly, diffusion and gravitional settling can affect the absolute abundances, but typically, leave relative abundances essentially unchanged, at least as predicted for the Sun (see e.g., \citealp{asp09} and references therein).}. Therefore, studying the detailed abundance pattern of solar-type stars may help us to re-construct the fundamental build-up of the Galaxy using chemical tagging \citep{fb02}. The goal of chemical tagging is to identify common origins for groups of stars based on their common, and unique, chemical compositions. A key assumption is that open clusters, which are the basic building blocks of the Galactic disc, should be chemically homogeneous. Most of previous studies (see e.g., \citealp{fb92,pau03,de06,de07,tin12,fri14}) have argued that open clusters are indeed chemically homogeneous. However, the observed abundance dispersions are typically $\approx$ 0.05 dex or larger, and can be attributed entirely to the measurement uncertainties. \citet{liu16b} discovered for the first time that the Hyades, a bench-mark open cluster, is chemically inhomogeneous at the 0.02 dex level, providing new constraints to the concept of chemical tagging. It is essential to conduct high-precision abundance analysis of other open clusters to further understand whether other open clusters are also chemically inhomogeneous.

In this paper, we perform a strictly line-by-line differential abundance analysis of two solar twins in M67 (M67-1194 and M67-1315). The aims of our analysis include: (a) examining the scenarios related to the peculiar chemical pattern of the Sun and the M67 solar twins; (b) testing the possible difference in chemical abundances of these two M67 stars, which could provide us with new clues as regards to the concept of chemical tagging.

\section{Observations and data reduction}

\begin{figure}
\centering
\includegraphics[width=\columnwidth]{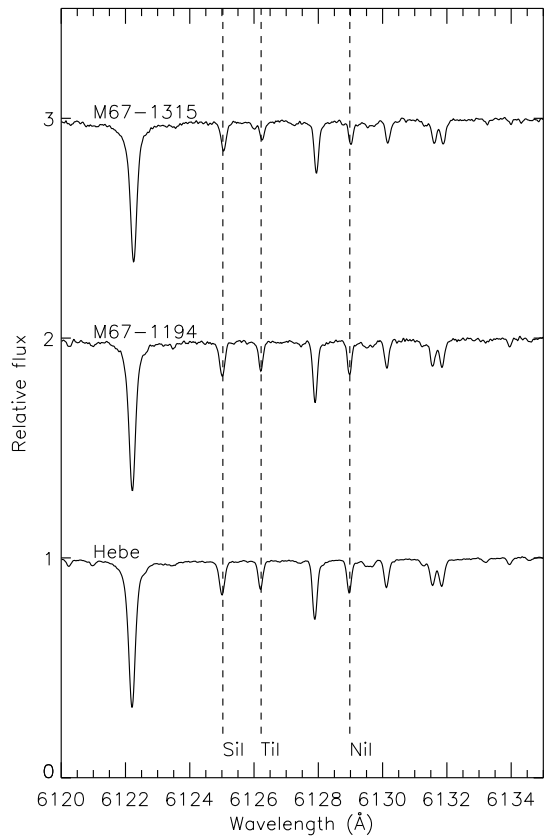}
\caption{A portion of the spectra for the programme stars. A few atomic lines (Si\,{\sc i}, Ti\,{\sc i}, Ni\,{\sc i}) used in our analysis in this region are marked by the dashed lines.}
\label{fig0}
\end{figure}

We observed two solar-type members in M67: M67-1194 and M67-1315. One of which (M67-1194) has been studied previously by O11. We obtained high resolution (R = $\lambda/\Delta\lambda$ = 50,000), high signal-to-noise ratio (SNR) spectra with the 0.86" slit of the High Resolution Echelle Spectrometer (HIRES, \citealp{vog94}) on the 10 m Keck I telescope during January 24 - 25, 2015. The spectral wavelength coverage is nearly complete from 400 to 840 nm. We obtained 12 spectra for M67-1194 and 11 spectra for M67-1315 with a total exposure time of 25800s for each star. The Keck-MAKEE pipeline was used for standard echelle spectra reduction including bias subtraction, flat-fielding, scattered-light subtraction, spectra extraction and wavelength calibration. We normalized and co-added the spectra with IRAF\footnote{IRAF is distributed by the National Optical Astronomy Observatory, which is operated by Association of Universities for Research in Astronomy, Inc., under cooperative agreement with National Science Foundation.}. The individual frames of each star were combined into a single spectrum with SNR $\approx$ 270 per pixel in most wavelength regions. A solar spectrum with even higher SNR ($\approx$ 450 per pixel) was obtained by observing the asteroid Hebe with the same instrumental configuration and reduced using the same method. A portion of the reduced spectra for the programme stars is shown in Figure \ref{fig0}.

We note that both programme stars were identified as solar twins (i.e., stars with indistinguishable stellar atmospheric parameters and metallicities to the Sun) in M67 by \citet{pas08}, based on their proper motion, radial velocities and colours. Constraints on binarity for our two sample stars have been provided by the radial velocity planet search on M67 stars by \citet{pas08} and \citet{pas12} in combination with our observations. The star M67-1194 seems stable, while a clear radial velocity trend was detected for M67-1315, suggesting that this solar twin is a binary. Our heliocentric radial velocity measurement of M67-1315 (36.4 $\pm$ 1.0 km\,s$^{-1}$) is also significantly different from the value 32.55 $\pm$ 0.34 km\,s$^{-1}$, as reported in \citet{pas08}, supporting that M67-1315 has a companion.
 
The spectral line-list of 28 elements (C, O, Na, Mg, Al, Si, S, Ca, Sc, Ti, V, Cr, Mn, Fe, Co, Ni, Cu, Zn, Sr, Y, Zr, Ba, La, Ce, Pr, Nd, Sm, Eu) employed in our analysis was adopted mainly from \citet{sco15a,sco15b} and \citet{gre15}, and complemented with additional unblended lines from \citet{ben05} and \citet{mel14}. Equivalent widths (EWs) were measured using the ARES code \citep{sou07} for a few species with a large number of spectral lines (i.e., Fe\,{\sc i}, Si, Ti\,{\sc i}, V, Cr\,{\sc i}, Ni), while EWs of other species with fewer spectral lines (i.e., C, O, Na, Mg, Al, S, Ca, Sc, Ti\,{\sc ii}, Cr\,{\sc ii}, Mn, Fe\,{\sc ii}, Co, Cu, Zn, Sr, Y, Zr, Ba, La, Ce, Pr, Nd, Sm, Eu) were measured interactively using the \textit{splot} task in IRAF. Strong lines with EW $\ge$ 110 m\AA\ were excluded from the analysis to limit the effects of saturation. The atomic line data, as well as the EW measurements, adopted for our analysis are listed in Table A1. 

\section{Stellar atmospheric parameters}

We performed a 1D, local thermodynamic equilibrium (LTE) chemical abundance analysis using the 2013 version of MOOG \citep{sne73,sob11} with the ODFNEW grid of Kurucz model atmospheres \citep{cas03}. Stellar atmospheric parameters (i.e., effective temperature T$_{\rm eff}$, surface gravity $\log g$, microturbulent velocity $\xi_{\rm t}$, and metallicity [Fe/H]) were obtained by forcing excitation and ionization balance of Fe\,{\sc i} and Fe\,{\sc ii} lines on a strictly line-by-line basis relative to the Sun. The adopted parameters for the Sun are: T$_{\rm eff} = 5777$\,K, $\log g$ = 4.44 [cm\,s$^{-2}$], $\xi_{\rm t}$ = 1.00 km\,s$^{-1}$, and [Fe/H] = 0.00. The stellar parameters of M67-1194 and M67-1315 were established separately using an automatic grid searching technique described by \citet{liu14}. In general, the best combination of T$_{\rm eff}$, $\log g$, $\xi_{\rm t}$, and [Fe/H], minimizing the slopes in [Fe\,{\sc i}/H] versus lower excitation potential (LEP) and reduced EW ($\log$\,(EW/$\lambda$) as well as the difference between [Fe\,{\sc i}/H] and [Fe\,{\sc ii}/H], was obtained from a successively refined grid of stellar atmospheric models. The final solution was obtained when the grid step-size decreased to $\Delta$T$_{\rm eff}$ = 1 K, $\Delta \log g$ = 0.01 [cm\,s$^{-2}$] and $\Delta$$\xi_{\rm t}$ = 0.01 km\,s$^{-1}$. We also require the derived average [Fe/H] to be consistent with the adopted model atmospheric value. Lines whose abundances departed from the average by $> 2.5\,\sigma$ were clipped.

\begin{figure}
\centering
\includegraphics[width=\columnwidth]{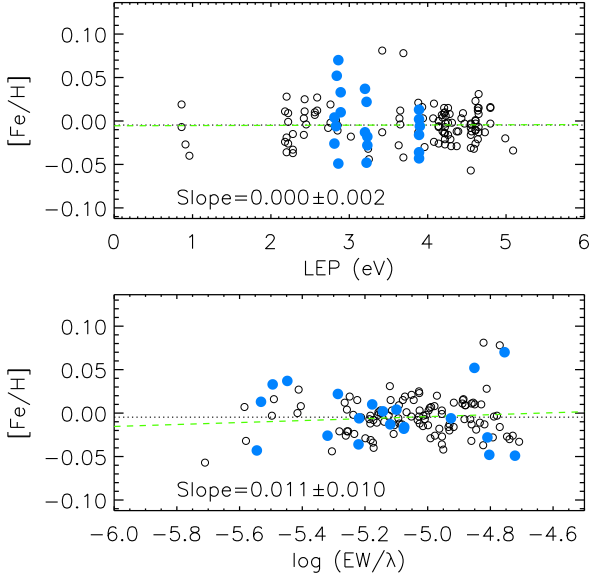}
\caption{Top panel: [Fe/H] of M67-1194 derived on a line-by-line basis with respect to the Sun as a function of LEP; open circles and blue filled circles represent Fe\,{\sc i} and Fe\,{\sc ii} lines, respectively. The black dotted line shows the location of mean [Fe/H], the green dashed line represents the linear least-squares fit to the data. Bottom panel: same as in the top panel but as a function of reduced EW.}
\label{fig1}
\end{figure}

Figure \ref{fig1} shows an example of determining the stellar atmospheric parameters of M67-1194. The adopted stellar parameters satisfy the excitation and ionization balance in a differential sense. The uncertainties correspond to $\Delta$T$_{\rm eff}$ $\approx$ 13 K, $\Delta\xi_{\rm t}$ $\approx$ 0.02 km\,s$^{-1}$ and $\Delta\log g$ $\approx$ 0.02 [cm\,s$^{-2}$]. These values are simply associated with the slopes of [Fe/H] versus LEP, [Fe/H] versus reduced EW and the difference between [Fe\,{\sc i}/H] and [Fe\,{\sc ii}/H], respectively.

The final adopted stellar atmospheric parameters of M67-1194 and M67-1315 with the Sun and its parameters as a reference point, are listed in Table \ref{t:para}. The adopted uncertainties in the stellar parameters were calculated using the method described by \citet{eps10} and \citet{ben14}, which accounts for the co-variances between changes in the stellar parameters and the differential iron abundances. High precision was achieved due to the high quality spectra and the strictly line-by-line differential method, which greatly reduces the systematic errors from atomic line data, shortcomings in the 1D LTE modelling of the stellar atmospheres and spectral line formation (see e.g., \citealp{asp05,asp09}). We find that our results for M67-1194 are similar to the values determined by O11: $\Delta$T$_{\rm eff}$ = 6 K, $\Delta\log g$ = 0.02 [cm\,s$^{-2}$], $\Delta$[Fe/H] = $-$0.028 (our work $-$ O11 work), while the uncertainties on our results are smaller by a factor of 1.5 - 2 due to the higher SNR of our combined spectrum\footnote{The SNR of combined spectrum for M67-1194 is $\approx$ 160 per pixel in O11 work compared to $\approx$ 270 per pixel in this work.}. We note that M67-1315, is warmer than M67-1194 by $\approx$ 150 K as well as slightly more metal-poor than M67-1194 by $\approx$ 0.05 dex.

\begin{table}
\caption{Adopted solar parameters and derived stellar parameters of M67-1194 and M67-1315 (relative to the Sun).}
\label{t:para}
\begin{tabular*}{\columnwidth}{lcccc}
\hline
Object & T$_{\rm eff}$ & $\log g$ & $\xi_{\rm t}$ & [Fe/H] \\
 & (K) & [cm\,s$^{-2}$] & (km\,s$^{-1}$) & \\
\hline
Hebe (Sun) & 5777 & 4.44 & 1.00 & 0.00 \\
\hline
M67-1194 & 5786$\pm$13 & 4.46$\pm$0.02 & 1.04$\pm$0.02 & $-$0.005$\pm$0.010 \\
M67-1315 & 5933$\pm$23 & 4.47$\pm$0.05 & 1.05$\pm$0.04 & $-$0.061$\pm$0.014 \\
\hline
\end{tabular*}
\end{table}

\section{Results and discussions}

\subsection{Elemental abundances of M67-1194 and M67-1315}

Having established the stellar atmospheric parameters of M67-1194 and M67-1315, we then derived chemical abundances for 27 elements in addition to Fe (C, O, Na, Mg, Al, Si, S, Ca, Sc, Ti, V, Cr, Mn, Co, Ni, Cu, Zn, Sr, Y, Zr, Ba, La, Ce, Pr, Nd, Sm, and Eu) based on the spectral lines and EW measurements listed in Table A1. We derived line-by-line differential elemental abundances [X/Fe] of M67-1194 and M67-1315 relative to the Sun. In addition, we also calculated line-by-line differential abundances $\Delta$[X/Fe] of M67-1315 relative to M67-1194, with the same setting of stellar parameters. We note that we consider an element with different ionization stages as two species in our analysis, rather than combining their elemental abundances together. We took hyperfine structure (HFS) into account for 9 elements (Sc, V, Mn, Co, Cu, Y, Ba, La, Pr) and calculated corrections strictly line-by-line. The HFS data were taken from \citet{kur95}. We note that the corrections are too small ($\le$ 0.005 dex) to affect our results. Non-LTE (NLTE) effects in the differential abundance analysis between very similar stars as here should be very small \citep{mel12,mon13}. However, we still applied differential NLTE corrections for the O\,{\sc i} triplet \citep{ama16}. The differential NLTE corrections relative to the Sun on [O/Fe] are $\approx$ 0.00 dex and $-$0.02 dex for M67-1194 and M67-1315, respectively. Table \ref{t:abun} lists the inferred differential chemical abundances for these 27 elements (29 species).

The errors in the elemental abundances were calculated following the method described by \citet{eps10}: the standard errors in the mean abundances, as derived from the different spectral lines, were added in quadrature to the errors introduced by the uncertainties in the stellar atmospheric parameters. For elements where only one spectral line was measured (i.e., Zr and La), we estimate the uncertainties by taking into consideration errors due to SNR, continuum setting and the stellar parameters. The quadratic sum of the three uncertainties sources give the errors for these two elements. The inferred errors on differential elemental abundances are listed in Table \ref{t:abun}. Most derived elemental abundances have uncertainties $\le$ 0.02 dex (for M67-1194) and $\le$ 0.03 dex (for M67-1315), which further underscores the advantages of a strictly differential abundance analysis on similar stars. When considering all species, the average uncertainty is 0.015 $\pm$ 0.001 ($\sigma$ = 0.004) for M67-1194 and 0.019 $\pm$ 0.001 ($\sigma$ = 0.005) for M67-1315, relative to the Sun, respectively. The average uncertainty is 0.021 $\pm$ 0.001 ($\sigma$ = 0.006) for line-by-line differential abundances when comparing M67-1315 to M67-1194. We note that the errors in stellar parameters and elemental abundances for M67-1315 are slightly higher than for M67-1194. The possible reason is that M67-1194 is essentially identical to the Sun in terms of T$_{\rm eff}$ and [Fe/H], while M67-1315 is slightly warmer by $\approx$ 150 K and more metal-poor by $\approx$ 0.06 dex.

\begin{table*}
\caption{Differential chemical abundances of M67-1194 and M67-1315 for 27 elements (29 species).}
\label{t:abun}
\begin{tabular}{@{}crrrr@{}}
\hline
 & \multicolumn{1}{r}{$\Delta$[X/Fe]} & \multicolumn{1}{r}{$\Delta$[X/Fe]} & \multicolumn{1}{r}{$\Delta$[X/Fe]} & \multicolumn{1}{r}{$\Delta$[X/H]} \\
Species & M67-1194$^a$ & M67-1315$^b$ & M67-1315$^c$ & M67-1315$^d$\\
\hline
 C\,{\sc i}  &  0.021 $\pm$ 0.015 &  0.046 $\pm$ 0.023 &  0.025 $\pm$ 0.029 & $-$0.031 $\pm$ 0.022 \\
 O\,{\sc i}  & $-$0.003 $\pm$ 0.016 & $-$0.023 $\pm$ 0.027 & $-$0.020 $\pm$ 0.032 & $-$0.076 $\pm$ 0.021 \\
Na\,{\sc i}  & $-$0.005 $\pm$ 0.014 & $-$0.015 $\pm$ 0.020 & $-$0.009 $\pm$ 0.023 & $-$0.065 $\pm$ 0.021 \\
Mg\,{\sc i}  & $-$0.003 $\pm$ 0.011 & $-$0.014 $\pm$ 0.028 & $-$0.011 $\pm$ 0.027 & $-$0.067 $\pm$ 0.025 \\
Al\,{\sc i}  &  0.011 $\pm$ 0.018 &  0.036 $\pm$ 0.019 &  0.025 $\pm$ 0.012 & $-$0.031 $\pm$ 0.010 \\
Si\,{\sc i}  & $-$0.001 $\pm$ 0.013 & $-$0.016 $\pm$ 0.014 & $-$0.015 $\pm$ 0.015 & $-$0.071 $\pm$ 0.009 \\
 S\,{\sc i}  &  0.001 $\pm$ 0.020 &  0.037 $\pm$ 0.021 &  0.036 $\pm$ 0.028 & $-$0.020 $\pm$ 0.023 \\
Ca\,{\sc i}  &  0.010 $\pm$ 0.010 &  0.024 $\pm$ 0.016 &  0.014 $\pm$ 0.016 & $-$0.042 $\pm$ 0.015 \\
Sc\,{\sc ii} & $-$0.003 $\pm$ 0.013 & $-$0.012 $\pm$ 0.018 & $-$0.009 $\pm$ 0.018 & $-$0.065 $\pm$ 0.023 \\
Ti\,{\sc i}  &  0.009 $\pm$ 0.011 &  0.036 $\pm$ 0.015 &  0.027 $\pm$ 0.017 & $-$0.029 $\pm$ 0.023 \\
Ti\,{\sc ii} & $-$0.005 $\pm$ 0.014 &  0.023 $\pm$ 0.019 &  0.027 $\pm$ 0.021 & $-$0.029 $\pm$ 0.026 \\
 V\,{\sc i}  &  0.027 $\pm$ 0.015 &  0.002 $\pm$ 0.022 & $-$0.025 $\pm$ 0.023 & $-$0.081 $\pm$ 0.029 \\
Cr\,{\sc i}  &  0.012 $\pm$ 0.012 &  0.004 $\pm$ 0.014 & $-$0.008 $\pm$ 0.017 & $-$0.064 $\pm$ 0.020 \\
Cr\,{\sc ii} & $-$0.011 $\pm$ 0.017 & $-$0.008 $\pm$ 0.017 &  0.004 $\pm$ 0.018 & $-$0.052 $\pm$ 0.019 \\
Mn\,{\sc i}  &  0.006 $\pm$ 0.012 & $-$0.006 $\pm$ 0.015 & $-$0.012 $\pm$ 0.018 & $-$0.068 $\pm$ 0.019 \\
Co\,{\sc i}  &  0.039 $\pm$ 0.015 & 0.026 $\pm$ 0.022 & $-$0.012 $\pm$ 0.021 & $-$0.068 $\pm$ 0.025 \\
Ni\,{\sc i}  & $-$0.008 $\pm$ 0.009 & $-$0.018 $\pm$ 0.012 & $-$0.010 $\pm$ 0.012 & $-$0.066 $\pm$ 0.016 \\
Cu\,{\sc i}  & $-$0.028 $\pm$ 0.008 & $-$0.010 $\pm$ 0.016 &  0.018 $\pm$ 0.015 & $-$0.038 $\pm$ 0.019 \\
Zn\,{\sc i}  & $-$0.025 $\pm$ 0.015 & $-$0.026 $\pm$ 0.015 & $-$0.001 $\pm$ 0.013 & $-$0.057 $\pm$ 0.011 \\
Sr\,{\sc i}  & $-$0.008 $\pm$ 0.022 &  0.068 $\pm$ 0.024 &  0.076 $\pm$ 0.024 &  0.020 $\pm$ 0.030 \\
 Y\,{\sc ii} &  0.016 $\pm$ 0.020 &  0.013 $\pm$ 0.019 & $-$0.003 $\pm$ 0.028 & $-$0.059 $\pm$ 0.033 \\
Zr\,{\sc ii} &  0.035 $\pm$ 0.023 &  0.093 $\pm$ 0.026 &  0.058 $\pm$ 0.027 &  0.002 $\pm$ 0.033 \\
Ba\,{\sc ii} &  0.008 $\pm$ 0.011 &  0.009 $\pm$ 0.015 &  0.001 $\pm$ 0.018 & $-$0.055 $\pm$ 0.021 \\
La\,{\sc ii} & $-$0.009 $\pm$ 0.024 &  0.057 $\pm$ 0.030 &  0.066 $\pm$ 0.032 &  0.010 $\pm$ 0.038 \\
Ce\,{\sc ii} &  0.016 $\pm$ 0.020 &  0.074 $\pm$ 0.025 &  0.058 $\pm$ 0.027 &  0.002 $\pm$ 0.033 \\
Pr\,{\sc ii} &  0.021 $\pm$ 0.018 &  0.055 $\pm$ 0.021 &  0.034 $\pm$ 0.024 & $-$0.022 $\pm$ 0.031 \\
Nd\,{\sc ii} &  0.017 $\pm$ 0.017 &  0.090 $\pm$ 0.022 &  0.072 $\pm$ 0.025 &  0.016 $\pm$ 0.033 \\
Sm\,{\sc ii} &  0.016 $\pm$ 0.016 &  0.071 $\pm$ 0.021 &  0.055 $\pm$ 0.023 & $-$0.001 $\pm$ 0.029 \\
Eu\,{\sc ii} & $-$0.011 $\pm$ 0.015 &  0.044 $\pm$ 0.018 &  0.055 $\pm$ 0.018 & $-$0.001 $\pm$ 0.026 \\
\hline
\end{tabular}
\\
$^a$ Differential abundance ratios $\Delta$[X/Fe] of M67-1194 relative to the Sun.\\
$^b$ Differential abundance ratios $\Delta$[X/Fe] of M67-1315 relative to the Sun.\\
$^c$ Differential abundance ratios $\Delta$[X/Fe] of M67-1315 relative to M67-1194.\\
$^d$ Differential abundances $\Delta$[X/H] of M67-1315 relative to M67-1194.\\
\end{table*}

\subsection{Elemental abundances versus T$_{\rm cond}$}

As noted in the Introduction, \citet{mel09} carried out a high-precision differential abundance analysis of the Sun and solar twins. They found that the Sun shows a depletion of refractory elements relative to volatile elements when compared to the majority ($\approx$ 85\%) of solar twins. Their results revealed a clear correlation between differential abundances and T$_{\rm cond}$. They tentatively attributed the peculiar chemical pattern of the Sun to the formation of terrestrial planets in the early solar system. Since both our M67 programme stars are solar twins/analogs, it is interesting to examine the T$_{\rm cond}$ trends (i.e., differential chemical abundances as a function of T$_{\rm cond}$). Detailed analysis of the T$_{\rm cond}$ trends can thus provide constraints on the scenarios related to the terrestrial planet formation (e.g., \citealp{mel09}) or dust-cleansing of cluster environments (e.g., O11 and O14). 

\begin{figure}
\centering
\includegraphics[width=\columnwidth]{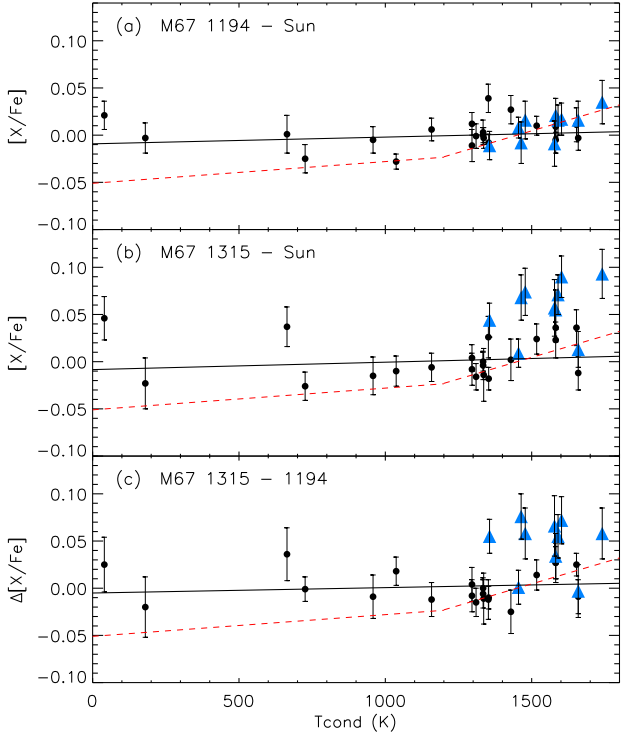}
\caption{Top panel: [X/Fe] of M67-1194 relative to the Sun as a function of T$_{\rm cond}$; black filled circles represent species with atomic number Z $\le$ 30 and blue filled triangles represent neutron-capture elements, respectively. The black solid line represents the linear least-squares fit to elemental abundances with Z $\le$ 30. The red dashed line represents the fitting of $<$solar twins $-$ Sun$>$ from \citet{mel09}. Middle panel: similar as in the top panel but for [X/Fe] of M67-1315 relative to the Sun. Bottom panel: similar as in the top panel but for $\Delta$[X/Fe] of M67-1315 relative to M67-1194.}
\label{fig2}
\end{figure}

Figure \ref{fig2} shows the differential elemental abundances ([X/Fe]) of M67-1194 and M67-1315 relative to the Sun, as well as $\Delta$[X/Fe] of M67-1315 relative to M67-1194 as a function of T$_{\rm cond}$\footnote{T$_{\rm cond}$ of each element was taken from \citet{lod03}.} in the top, middle, and bottom panel, respectively. From our results, M67-1194 is indistinguishable from the Sun in terms of its chemical composition. The average abundance difference between M67-1194 and the Sun is $<$[X/Fe]$>$ = 0.005 $\pm$ 0.003 ($\sigma$ = 0.015). The dispersion is consistent with the average error stated earlier (0.015 $\pm$ 0.001 dex). The slope of the linear least-squares fit to all the data points is (1.15 $\pm$ 0.69) $\times$ 10$^{-5}$ K$^{-1}$. As shown in the middle and bottom panel of Figure \ref{fig2}, the chemical pattern of M67-1315 is similar when compared to the Sun or to M67-1194. We note that M67-1315 shares similar elemental abundances to the Sun as well as M67-1194 for elements with atomic number Z $\le$ 30. The average abundance difference (Z $\le$ 30) between M67-1315 and the Sun is $<$[X/Fe]$>$ = 0.004 $\pm$ 0.005 ($\sigma$ = 0.022), where the dispersion is also comparable with the measurement errors stated before (0.019 $\pm$ 0.001 dex). The average abundance difference (Z $\le$ 30) between M67-1315 and M67-1194 is $<$$\Delta$[X/Fe]$>$ = 0.002 $\pm$ 0.004 ($\sigma$ = 0.018), and we note again that the dispersion is consistent with the average error (0.021 $\pm$ 0.001 dex). The slopes of the linear least-squares fits to lighter elements (Z $\le$ 30) for M67-1315 are (0.78 $\pm$ 1.05) $\times$ 10$^{-5}$ K$^{-1}$ and (0.57 $\pm$ 1.11) $\times$ 10$^{-5}$ K$^{-1}$, respectively.

\begin{figure}
\centering
\includegraphics[width=\columnwidth]{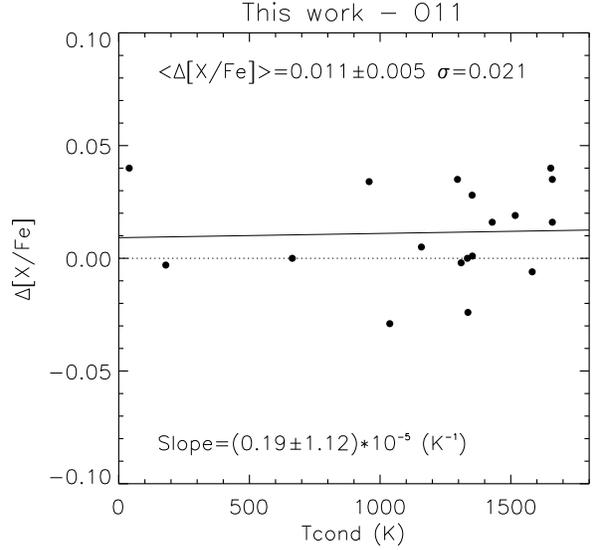}
\caption{Difference in abundances ($\Delta$[X/Fe]) between this work and O11 for M67-1194. The black solid line represents the linear least-squares fit to the comparison data.}
\label{fig3}
\end{figure}

O11 performed a differential abundance analysis of the common star M67-1194 relative to the Sun. They found that the chemical composition of M67-1194 is similar to the solar composition and proposed the possible scenario for their results is that the Sun was born in a dense environment similar to M67. We compared our abundances with that of O11 for elements in common. As seen in Figure \ref{fig3}, the average difference between our results and O11 results is $<$$\Delta$[X/Fe]$>$ = 0.011 $\pm$ 0.005 ($\sigma$ = 0.021). Our [X/Fe] ratios are slightly higher, mainly due to the difference in [Fe/H]: we derived [Fe/H] = $-$0.005 $\pm$ 0.010, while O11 determined [Fe/H] = $+$0.023 $\pm$ 0.015. 

Our results therefore indicate that both programme stars in M67 show similar chemical patterns to the "peculiar" solar chemical pattern for elements with Z $\le$ 30. As mentioned before, M67-1194 hosts a hot Jupiter with a period of 6.9 days and a minimum mass of 0.34 M$_{\rm Jup}$ \citep{bru14}. This is thus not in line with the planet formation scenario and might favour the dust-cleansing scenario proposed by O11 and O14. \citet{bru16} reported that the frequency of hot Jupiters in M67 is $\sim$ 5.6\% for single stars and $\sim$ 4.5\% for the full sample. Our sample consists of only two stars so more high quality spectra of M67 solar-type stars are needed to examine all the possible explanations and constraints on the rate of planets in this open cluster.

We note that the above results are shown based on $\Delta$[X/Fe], rather than $\Delta$[X/H]. As stated before, M67-1315 is more metal-poor than M67-1194 by $\approx$ 0.05 dex. When compared M67-1315 to M67-1194 in terms of $\Delta$[X/H], as listed in the last column, Table \ref{t:abun}, there will be an offset due to the higher overall metallicity of M67-1194. In this case, M67-1315 could be poor in lighter elements (Z $\le$ 30) and only normal in neutron-capture elements. No clear explanation could be made. We emphasize that our main results are presented based on $\Delta$[X/Fe].

\subsection{M67-1315: enrichment of abundances in neutron-capture elements}

As seen in Figure \ref{fig2}, the elemental abundances for M67-1194 are very similar to the solar values. The chemical abundance pattern is more complicated for M67-1315. We recall that the lighter elements (Z $\le$ 30) of M67-1315 show similar abundance behaviour with that of the Sun and M67-1194. However, the abundances of neutron-capture elements of M67-1315 are enriched by $\approx$ 0.05 dex relative to the Sun, except for Y and Ba. After taking into account the measurement uncertainties, the enrichment is significant at the 4.5\,$\sigma$ level. This complex chemical behaviour is unexpected and worth further exploration. In the following analysis, we focus on $\Delta$[X/Fe] of neutron-capture elements of M67-1315 relative to M67-1194. 

\begin{figure}
\centering
\includegraphics[width=\columnwidth]{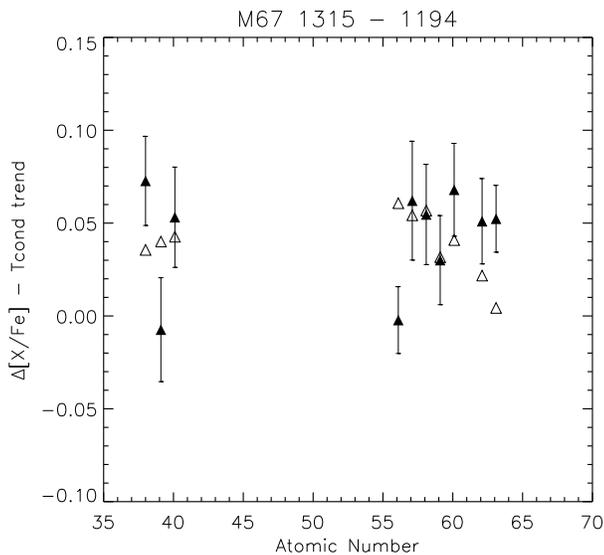}
\caption{Neutron-capture abundances of M67-1315 relative to M67-1194 after subtraction of T$_{\rm cond}$ trend. The filled triangles represent our results, while the open triangles are the $\Delta$[X/Fe]$_{\rm AGB}$ ratios.}
\label{fig4}
\end{figure}

We followed the method described by \citet{mel14}. At first we fit $\Delta$[X/Fe] versus T$_{\rm cond}$ for the elements with Z $\le$ 30 and obtain:
\begin{equation}
\Delta \rm [X/Fe] \,(Z \le 30) = -0.005 + 0.570 \times 10^{-5} \, T_{cond}
\end{equation}
The element-to-element scatter is 0.019 dex, which is similar to the average total error (0.021 dex) of our differential abundances, showing that the error bars in our analysis are realistic. Next, we subtract the above T$_{\rm cond}$ trend from the [X/Fe] abundances:
\begin{equation}
\Delta \rm [X/Fe]_T = \Delta[X/Fe] - (-0.005 + 0.570 \times 10^{-5} \, T_{cond})
\end{equation}
As mentioned before, M67-1315 likely has a companion. AGB stars are efficient producers of neutron-capture elements via the slow neutron capture process (e.g., \citealp{kl14}), so it is possible that the enhancement in neutron-capture elements for M67-1315 is due to pollution from the companion, when it was in its AGB phase (now should be a white dwarf). In order to estimate if an AGB companion star has polluted M67-1315, we tried different AGB models from \citet{kar16}, scaled to our results, to find the best match using $\chi^2$ as the metric. Finally, the best fit was obtained when using a 2 M$_\odot$ AGB star of solar metallicity and diluted those yields by mixing a small fraction of AGB ejecta with the convective zone material ($\sim$ 0.02 M$_\odot$) of solar composition \citep{asp09}. A dilution of 0.0092\% mass of AGB material matched the observed enhancement, as determined with minimum $\chi^2$ ($\chi^2_{\rm reduced}$ = 1.73). We plot differential abundances of neutron-capture elements of M67-1315 after subtraction of T$_{\rm cond}$ trend in Figure \ref{fig4}. The best model predicted ratios ($\Delta$[X/Fe]$_{\rm AGB}$) are shown by open triangles. As can be seen, a good match can not be achieved for all the neutron-capture elements, mainly because the two $s$-process elements Y and Ba are not enhanced in M67-1315.

\begin{figure}
\centering
\includegraphics[width=\columnwidth]{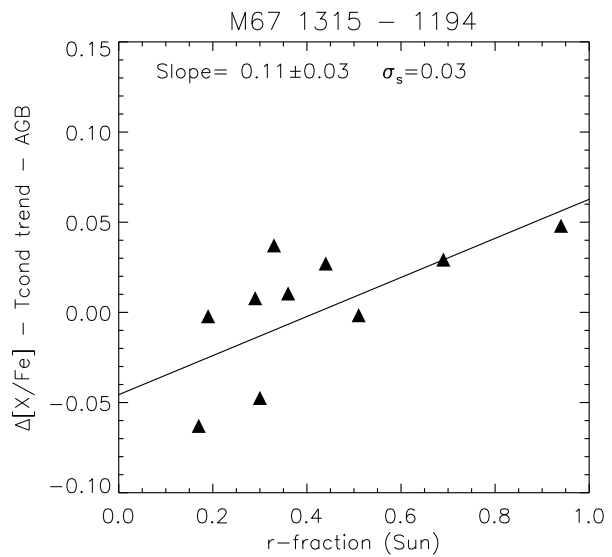}
\caption{Neutro-capture abundance residuals ($\Delta$[X/Fe]$_r$) of M67-1315 relative to M67-1194 versus the solar system \textit{r} fractions \citep{bis11,bis13}. The black solid line represents the linear least-squares fit to the data.}
\label{fig5}
\end{figure}

In order to investigate if the residual enrichment was due to the \textit{r}-process, we estimated the neutron-capture abundance enhancement in M67-1315 by subtracting the AGB contribution:
\begin{equation}
\Delta \rm [X/Fe]_\textit{r} = \Delta[X/Fe]_T - \Delta[X/Fe]_{AGB}
\end{equation}
and compared our results with the predicted enrichment based on the \textit{r}-process fractions in the solar system, using the \textit{s}-process fractions (\textit{s}$_{\rm bis}$) from \citet{bis11} and \citet{bis13}. We plot $\Delta$[X/Fe]$_r$ of neutron-capture elements of M67-1315 versus the solar \textit{r} fractions (\textit{r} = 1 $-$ \textit{s}$_{\rm bis}$) in Figure \ref{fig5}. It seems possible that the residuals, when plotted against \textit{r} fractions, could be due to the scaled solar \textit{r}-process. The best linear least-squares fit has a significance of 3.9\,$\sigma$ and the dispersion about the linear fit: 0.029 dex, is comparable to the average error (0.025 dex) for these neutron-capture elements. This comparison demonstrates that after both the T$_{\rm cond}$ trend and AGB pollutions are subtracted, most of the residual material can be possibly explained by the \textit{r}-process. Similar findings can be found for some solar twins (e.g., \citealp{mel14,yg16b}) and metal-poor stars (e.g., \citealp{fre07,sne08}). The fact that we find an \textit{r}-element enrichment without any significant enrichment of $\alpha$-elements is difficult to reconcile with a core-collapse supernova origin for the \textit{r}-process and more consistent with neutron star mergers (e.g., \citealp{arn07,gor11,wan14}). Table \ref{t:ncap} lists the neutron-capture enhancement in M67-1315 relative to M67-1194 and related information.

\begin{table}
\caption{Neutron-capture abundances in M67-1315, relative to M67-1194.}
\label{t:ncap}
\begin{tabular}{@{}ccrcrr@{}}
\hline
Z & Element & $\Delta$[X/Fe]$_{\rm T}$ & $\Delta$[X/Fe]$_{\rm AGB}^a$ & $\Delta$[X/Fe]$_r$ & \textit{s}$_{\rm bis}^b$ \\
\hline
38 & Sr &  0.073 & 0.036 &  0.037 & 0.67 \\
39 &  Y & $-$0.007 & 0.040 & $-$0.047 & 0.70 \\
40 & Zr &  0.053 & 0.043 &  0.011 & 0.64 \\
56 & Ba & $-$0.002 & 0.061 & $-$0.063 & 0.83 \\
57 & La &  0.062 & 0.054 &  0.008 & 0.71 \\
58 & Ce &  0.055 & 0.057 & $-$0.002 & 0.81 \\
59 & Pr &  0.030 & 0.032 & $-$0.002 & 0.49 \\
60 & Nd &  0.068 & 0.041 &  0.027 & 0.56 \\
62 & Sm &  0.051 & 0.022 &  0.029 & 0.31 \\
63 & Eu &  0.052 & 0.004 &  0.048 & 0.06 \\
\hline
\end{tabular}
\\ $^a$ $\Delta$[X/Fe]$_{\rm AGB}$ ratios are calculated using the AGB models provided in \citet{kar16}.
\\ $^b$ \textit{s}-process solar system fractions are taken from \citet{bis11,bis13}.\\
\end{table}

\subsection{Lithium abundances of M67-1194 and M67-1315}

LTE lithium abundances were determined by spectral synthesis with MOOG, as described in \citet{car16}. In short, we estimate first the stellar broadening using clean lines around 6000 \AA, and with the instrumental and stellar broadening fixed, we fit the lithium region using the line list of \citet{mel12}. NLTE corrections for lithium are computed using the grid of \citet{lind09}. Finally, the errors are estimated based on uncertainties in the continuum setting, the deviations of the observed profile relative to the synthetic spectrum, and the uncertainties introduced by the errors in the stellar parameters.

For M67-1194, the solar twin that most resemble the Sun in M67, we found A(Li)$_{\rm NLTE}$ = 1.36$^{+0.08}_{-0.07}$ dex. Our LTE value (1.32$^{+0.08}_{-0.07}$ dex) is in reasonable agreement with the previous results obtained in LTE by O11 and \citet{cas11}, A(Li)$_{\rm LTE}$ = 1.26 and $\le$ 1.3 dex, respectively. According to the Li-age correlation (e.g., \citealp{car16}), the Li abundances of solar twins in M67 should be somewhat higher than solar, because the cluster is probably somewhat younger than the Sun according to different previous studies using isochrones (e.g., \citealp{cas11} determined an age of 3.9 Gyr); also \citet{gia06} suggest an age determination younger than the Sun for M67 based on stellar activity. Indeed, the over-solar Li abundance of M67-1194 agrees with the Li-age correlation.

The Li abundance in M67-1315, that is hotter and likely more massive than the Sun, is A(Li)$_{\rm NLTE}$ = 1.90$^{+0.02}_{-0.03}$ dex (or 1.89 dex in LTE). Our result is in reasonable agreement with the abundance obtained by \citet{cas11}, A(Li)$_{\rm LTE}$ =  1.8 dex, based on a lower quality spectrum. Notice that the high Li abundance in M67-1315 agrees with the Li abundance of other solar analogs more massive than the Sun in M67, and also with the Li-mass relations predicted for this cluster \citep{pac12,cas16}.

\subsection{Implications for chemical tagging}

In the current view of Galactic archeology, star-forming aggregates imprint unique chemical signatures, which may be used to identify and track individual stars back to a common birth site, a concept normally referred to as chemical tagging \citep{fb02}. One of the pre-requisites for chemical tagging to be successful is that open clusters, which are surviving star-forming aggregates in the Galactic disc, should be chemically homogeneous \citep{bla10,bc15,de15,tin15}. Determining the level of chemical homogeneity in open clusters is thus of fundamental importance in the study of the evolution of star-forming clouds and that of the Galactic disc \citep{bov16}. \citet{liu16b} discovered for the first time, that the bench-mark open cluster Hyades is chemically inhomogeneous at the 0.02 dex level.

Within the similar precision level, our results for two solar-type M67 stars (M67-1194 and M67-1315) show peculiar chemical abundance patterns. M67-1194 is identical to the Sun in its chemical composition, while M67-1315 has a metallicity 0.06 dex lower than solar, albeit it shares similar [X/Fe] ratios with both the Sun and M67-1194 for Z $\le$ 30. For the neutron-capture elements, however, M67-1315 exhibits enhancements by a factor of 0.05 dex, except for Y and Ba. The somewhat different metallicity of M67-1315 may suggest that it is not a member, but all other indicators (proper motions, radial velocity, colours, magnitude) confirm indeed that the star is a member of M67. Therefore its peculiar metallicity most likely reflects intrinsic abundance variations in M67 stars. Assuming that M67-1194 is normal and M67-1315 is unusual,
it might be possible that M67 harbours a non-negligible fraction of stars that are enhanced in neutron-capture elements. In addition, the \textit{s}-process material could have come from a binary companion star, or from the birth cloud, while the \textit{r}-process material could only have come from the birth cloud of this star. Our results imply that the enhanced \textit{s}-process material of M67-1315 might come from its companion, rather than the birth cloud, while the enhanced \textit{r}-process material of M67-1315 could be related to neutron star mergers or supernovae events that happened during the early days of M67. \citet{hur05} have argued that M67 might have started with more than 20,000 stars. It is for the first time that clear evidence of star-to-star variations in neutron-capture elements in open clusters are reported with such a high precision chemical abundance analysis. We note that due to the small statistics, more high quality spectra of M67 solar-type stars are essential for further exploration of the implications for chemical tagging.

\section{Conclusions}

We performed a strictly line-by-line differential chemical abundance analysis of two solar twins in the M67 open cluster (M67-1194 and M67-1315). Stellar atmospheric parameters and chemical abundances for 29 elements were obtained with high precision using high resolution, very high SNR Keck/HIRES spectra. From our results, we confirm that M67-1194 is a solar-twin with stellar parameters and chemical composition identical to the Sun. We found that M67-1315 is a solar analog with effective temperature slightly warmer than the Sun by $\approx$ 150 K and more metal-poor than the Sun by $\approx$ 0.06 dex. The chemical abundance pattern of M67-1315 is complex: its metallicity is $\approx$ 0.05 dex lower than M67-1194 and the abundance ratios [X/Fe] of elements with Z $\le$ 30 are similar to both the Sun and M67-1194, while the abundance ratios of neutron-capture elements are enhanced by $\approx$ 0.05 dex, except for the $s$-process elements Y and Ba, which are not overabundant. The special neutron-capture elemental abundances could be attributed to AGB material as well as scaled-solar $r$-process material.

From our results, M67-1194 and M67-1315 show similar abundance behaviour with that of the Sun for lighter elements (Z $\le$ 30). This might favour the dust-cleansing scenario, proposed by O11 and O14, that the dust in the pre-solar nebula was radiatively cleansed by massive stars in a dense environment before the formation of the Sun and its planets, similar to what happened in the proto-cluster cloud of M67. The distinctive chemical abundances for the neutron-capture elements in M67-1315 and the lower metallicity of this star compared to M67-1194, indicate that the stars in M67 are likely not chemically homogeneous. Our results have implications for chemical tagging considering that two solar-type stars within an open cluster do have distinctive neutron-capture elemental abundances. However, we note that our sample only includes two stars. Therefore it is essential to have more high quality data in order to further investigate the peculiar chemical patterns of solar twins or solar analogs in M67 open cluster.

\section*{Acknowledgments}
This work has been supported by the Australian Research Council (grants FL110100012, FT140100554 and DP120100991). JM thanks support by FAPESP (2012/24392-2). The authors thank the ANU Time Allocation Committee for awarding observation time to this project. The authors wish to acknowledge the very significant cultural role and reverence that the summit of Mauna Kea has always had within the indigenous Hawaiian community. We are most fortunate to have the opportunity to conduct observations from this mountain.

\section*{SUPPLEMENTARY MATERIAL}

The following supplementary material is available for this article online:

Table A1. Atomic line data, as well as the EW measurements, adopted for our analysis.

\bsp

\label{lastpage}

\end{document}